\documentclass[11pt,a4paper]{article}
\usepackage[a4paper,hmargin=2.8cm,vmargin=3cm]{geometry}
\usepackage[utf8x]{inputenc}
\usepackage{amsmath,amsthm,amsfonts,amssymb}

\newtheorem*{theorem*}{Theorem}

\newcommand{\nielsRbb}{\mathbb{R}}
\newcommand{\nielsdi}{\mathrm{d}}
\newcommand{\nielsLs}{L^2_\mathrm{s}(\nielsRbb^{3N})}
\newcommand{\nielsnorm}[1]{\lVert #1 \rVert}
\newcommand{\nielstr}{\mathrm{tr}\,}
\newcommand{\nielsfock}{\mathcal{F}}
\newcommand{\nielsHcal}{\mathcal{H}}
\newcommand{\nielshc}{\mathrm{h.\,c.}}
\newcommand{\nielsNcal}{\mathcal{N}}
\newcommand{\nielsLcal}{\mathcal{L}}

\begin{document}

\title{Deriving the Gross-Pitaevskii equation}

\author{Niels Benedikter \smallskip \\
Institute of Applied Mathematics, University of Bonn\\
Endenicher Allee 60, 53115 Bonn, Germany\\
E-mail: niels.benedikter@tofoq.de}

\maketitle

\begin{abstract}
In experiments, Bose-Einstein condensates are prepared by cooling a dilute Bose gas in a trap. After  the phase transition has been reached, the trap is switched off and the evolution of the condensate observed. The evolution is macroscopically described by the Gross-Pitaevskii equation. On the microscopic level, the dynamics of Bose gases are described by the $N$-body Schr\"odinger equation. We review our article \cite{BdS} in which we construct a class of initial data in Fock space which are energetically close to the ground state and prove that their evolution approximately follows the Gross-Pitaevskii equation. The key idea is to model two-particle correlations with a Bogoliubov transformation.

\emph{Keywords: Bose-Einstein condensate; dilute Bose gas; Gross-Pitaevskii equation; correlations; many-body systems; Bogoliubov transformations; squeezed coherent states.}
\end{abstract}

\section{Bose-Einstein condensates in the Gross-Pitaevskii regime}
In dilute gases of bosonic particles, at very low temperatures a phase transition occurs and a macroscopic number of particles occupies the same one-particle state. Bose and Einstein theoretically predicted this state of matter in 1924, considering non-interacting bosons. The experimental confirmation took 70 years and was rewarded with a Nobel prize in 2001, and the theoretical study of interacting Bose-Einstein condensates still poses challenging problems.

\smallskip

A dilute gas of $N$ bosons can be described with the Hamilton operator
\[H^U_N = \sum_{j=1}^N \left(-\Delta_j + U(x_j)\right) + \sum_{i < j}^N N^2 V(N(x_i-x_j))\]
acting on the Hilbert space $\nielsLs$, the subspace of $L^2(\nielsRbb^{3N})$ symmetric with respect to permutation of the $N$ particles. The external potential $U$ models the trap that confines the particles. The interaction potential $V$ is  repulsive ($V\geq 0$) and spherically symmetric; the Gross-Pitaevskii scaling $N^2V(N.)$ models rare but strong collisions. We are interested in large $N$ (in experiments, $N \gg 10^3$).

A central role is played by the solution $f$ to the zero-energy scattering equation
\[\Big(-\Delta+\frac{1}{2}V\Big)f=0, \ \ \text{with boundary condition } f(x) \to 1 \ \ (\lvert x\rvert \to \infty).\]
Its solution has the form $f(x) \simeq 1 - a_0/\lvert x\rvert$ for large $x$, where
$a_0 := (8\pi)^{-1} \int f V \nielsdi x$ is the scattering length of $V$. By scaling, $f(N.)$ solves the zero-energy scattering equation with scaled potential $N^2 V(N.)$.

%
It was proven \cite{LSY} that the ground state energy $E_N$ of the Hamiltonian $H^U_N$ satisfies
\begin{equation}
\label{nielsbeq:scattering}
\lim_{N \to \infty} \frac{E_N}{N} = \min_{\varphi \in L^2(\nielsRbb^3),\, \nielsnorm{\varphi}_2=1} \mathcal{E}_\mathrm{GP}(\varphi),
\end{equation}
with the Gross-Pitaevskii energy functional
\begin{equation}\label{nielsbeq:gpfunctional}
\mathcal{E}_\mathrm{GP}(\varphi) := \int \nielsdi x \left( \lvert \nabla \varphi \rvert^2 + U \lvert \varphi\rvert^2 + 4\pi a_0 \lvert \varphi\rvert^4 \right).   
\end{equation}
The ground state $\psi^\mathrm{gs}_N$ exhibits \cite{LS} complete Bose-Einstein condensation, in the sense
\[\gamma^{(1)}_{\psi^\mathrm{gs}_N} \xrightarrow{\ \ \text{in trace norm}\ \ } \lvert \varphi_\mathrm{GP} \rangle\langle \varphi_\mathrm{GP} \rvert \quad (N \to \infty),\]
where $\lvert \varphi_\mathrm{GP} \rangle\langle \varphi_\mathrm{GP} \rvert$ is the projection on the  minimizer $\varphi_\mathrm{GP}$ of the Gross-Pitaevskii functional \eqref{nielsbeq:gpfunctional}, and $\gamma^{(1)}_{\psi^\mathrm{gs}_N}$ is the one-particle reduced density matrix associated with $\psi^\mathrm{gs}_N$, i.\,e.\ the trace-class operator on $L^2(\nielsRbb^3)$ defined through the integral kernel
\[\gamma^{(1)}_{\psi^\mathrm{gs}_N}(x;y) := \int \nielsdi x_2 \ldots \nielsdi x_N\, \psi^\mathrm{gs}_N(x,x_2,\ldots,x_N)\,\overline{\psi^\mathrm{gs}_N}(y,x_2,\ldots,x_N).\]
Since we generally assume $\nielsnorm{\psi}_2 = 1$, we have $\nielstr \gamma^{(1)}_\psi =1$.

When the traps are switched off ($U=0$), the system starts to evolve, following the Schr\"odinger equation
\[i \partial_t \psi_{N,t} = H^0_N \psi_{N,t}, \quad \psi_{N,0} = \psi^\mathrm{gs}_N.\] 
It was proven \cite{ESY2,Pickl} that $\gamma^{(1)}_{\psi_{N,t}} \to \lvert \varphi_t\rangle\langle \varphi_t\rvert$ as $N \to \infty$ (for any fixed $t>0$), where $\varphi_t$ is the solution to the non-linear Gross-Pitaevskii equation (here with initial data $\varphi = \varphi_\mathrm{GP}$) 
\begin{equation}\label{nielsbeq:GP}
i\partial_t \varphi_t = -\Delta \varphi_t + 8\pi a_0\lvert \varphi_t\rvert^2 \varphi_t,\quad \varphi_0 = \varphi.
\end{equation}

In our analysis we generalize the system to Fock space, with the advantage that we can use initial data that are superpositions of states with different numbers of particles.
We introduce the bosonic Fock space $\nielsfock := \bigoplus_{j=0}^\infty L^2_\mathrm{s}(\nielsRbb^{3j})$ and creation/annihilation operators (more precisely operator-valued distributions) $a^*_x$, $a_x$,
which create/annihilate a particle at $x \in \nielsRbb^3$. They satisfy the canonical commutation relations $[a_x,a^*_y] = \delta(x-y)$, $[a^*_x,a^*_y] = 0 = [a_x,a_y]$. We introduce the number of particles operator $\nielsNcal = \int \nielsdi x\, a^*_x a_x$ and the vacuum vector $\Omega = (1,0,\ldots) \in \nielsfock$. On Fock space $\nielsfock$ we define the Hamiltonian
\begin{equation}
\label{nielsbeq:Hcal}
\nielsHcal_N := \int \nielsdi x\, \nabla_x a^*_x \nabla_x a_x + \frac{1}{2N}\int \nielsdi x\nielsdi y\, N^3V(N(x-y)) a^*_x a^*_y a_y a_x. 
\end{equation}
The restriction of $\nielsHcal_N$ to $\nielsLs$ coincides with the Hamiltonian $H^0_N$.

For $g \in L^2(\nielsRbb^3)$ we define the Weyl operator
\[W(g) := \exp\left(\int \nielsdi x\,a^*_x g(x) - \nielshc\right),\]
and for integral kernels $k \in L^2(\nielsRbb^3 \times\nielsRbb^3)$ we introduce the Bogoliubov transformation
\begin{equation}
\label{nielsbeq:bogoliubov}
  T(k) := \exp\left(\frac{1}{2} \int \nielsdi x\nielsdi y\, k(x;y) a^*_x a^*_y - \nielshc \right).  
\end{equation}
The Weyl operators have the important property of shifting the operators,
\begin{equation}\label{nielsbeq:weyl}
   W^*(g) a^*_x W(g) = a^*_x + \bar g(x),\quad W^*(g) a_x W(g) = a_x + g(x),
\end{equation}
whereas the Bogoliubov transformation $T(k)$ acts by
\begin{equation}
 \label{nielsbeq:bogtrafo}
T^*(k) a^*_x T(k) = \int \nielsdi y\left( a^*_y \cosh(k)(y;x)  + a_y \sinh(k)(y;x)\right).
 \end{equation}
Here $\cosh(k)(y;x)$ and $\sinh(k)(y;x)$ are the integral kernels defined by the power series (in $k$) of the hyperbolic cosine/sine, with product in the sense of operators.

We will use a Weyl operator to generate a condensate. (The coherent state $W(g)\Omega$ describes a condensate with approximately $\nielsnorm{g}^2_2$ particles in the one-particle state $g/\|g\|_2$.) A Bogoliubov transformation, on the other hand, is used to implement correlations among the particles. Our main result \cite{BdS} is the following theorem.
\begin{theorem*}
Let $V \geq 0$ and $V \in L^1\!\cap\!L^3\!\left(\nielsRbb^3,(1+\lvert x\rvert^6)\nielsdi x\right)$. Let $\varphi \in H^4(\nielsRbb^3)$ with $\nielsnorm{\varphi}_{2} = 1$. Let $k_0(x;y) := -N \left(1-f\left(N\left(x-y\right)\right)\right) \varphi (x) \varphi(y)$. Let $\chi \in \nielsfock$, possibly depending on $N$ but s.\,t.\ $\langle \chi,\left(\nielsNcal+1 + \nielsNcal^2/N + \nielsHcal_N\right)\chi\rangle$ is bounded uniform in $N$. We consider $\psi_{N,t} := e^{-i\nielsHcal_N t}W(\sqrt{N} \varphi ) T(k_0 ) \chi$, the solution to the Schr\"odinger equation in Fock space, i.e.\ $i\partial_t \psi_{N,t} = \nielsHcal_N \psi_{N,t}$. Then there exist constants $C, c > 0$ s.\,t.\
\[\nielstr \left\lvert \gamma^{(1)}_{\psi_{N,t}} - \lvert \varphi_t\rangle\langle \varphi_t\rvert \right \rvert \leq \frac{C}{\sqrt{N}}\exp(c \exp(c \lvert t\rvert)),\]
where $\varphi_t$ solves the Gross-Pitaevskii equation \eqref{nielsbeq:GP} with initial data $\varphi_0 = \varphi$.
\end{theorem*}
The vector $\chi$ in the initial data describes small deviations from the squeezed coherent state $W(\sqrt{N} \varphi ) T(k_0 ) \Omega$. The correlation structure is inserted already in the initial data; our proof keeps this structure static, showing its approximate stability. Our initial data arises naturally as an approximation to the ground state since
\[\langle W(\sqrt{N} \varphi ) T(k_0 ) \chi,\big(\nielsHcal_N + \int\!\nielsdi x\,U(x) a^*_x a_x\big) W(\sqrt{N} \varphi ) T(k_0 ) \chi\rangle = N \mathcal{E}_\mathrm{GP}(\varphi) + \mathcal{O}(N^{1/2}).\]
In our article \cite{BdS} we also discuss initial data with exact number of particles.

\section{Strategy: Modeling correlations by a Bogoliubov transformation}
The approach is inspired by the method of coherent states \cite{RS}, developed for studying the mean-field regime.
However, coherent states cannot provide a good approximation in the Gross-Pitaevskii regime because they describe completely uncorrelated states. To take into account the correlations we use Bogoliubov transformations.

For technical reasons we will compare the solution of the many-body Schr\"odinger equation first to the solution of the modified Gross-Pitaevskii equation
\begin{equation}\label{nielsbeq:modGP}
i \partial_t \varphi^{(N)}_t = -\Delta \varphi^{(N)}_t + \big( N^3 f(N.) V(N.) \ast \lvert \varphi^{(N)}_t\rvert^2 \big)\varphi^{(N)}_t, \quad \varphi^{(N)}_0 = \varphi. 
\end{equation}
Since $N^3 f(N.) V(N.) \to 8\pi a_0 \delta$, it is easy to compare $\varphi^{(N)}_t$ with the solution $\varphi_t$ of the Gross-Pitaevskii equation \eqref{nielsbeq:GP}. With $f$ the solution to the zero-energy scattering equation, let
\[k_t(x;y) := - N \left(1-f\left(N\left(x-y\right)\right)\right)\varphi^{(N)}_t(x)\,\varphi^{(N)}_t(y).\]
We try to approximate the full evolution $\psi_{N,t} = e^{-i\nielsHcal_N t} W(\sqrt{N}\varphi) T(k_0) \chi$ with the (up to the small deviation) squeezed coherent state $W(\sqrt{N} \varphi^{(N)}_t) T(k_t)\chi$. 
Thus, inspired by \cite{RS}, we introduce the fluctuation dynamics
\[\mathcal{U}_N(t) := T^*(k_t) W^*(\sqrt{N} \varphi^{(N)}_t) e^{-i\nielsHcal_N t} W(\sqrt{N}\varphi) T(k_0).\]
%
%
%
We find the estimate
\begin{equation}
\label{nielsbeq:numberoffluct}
 \nielstr \left\lvert \gamma^{(1)}_{\psi_{N,t}} - \lvert \varphi^{(N)}_t \rangle \langle \varphi^{(N)}_t \rvert \right\rvert \leq \frac{C}{\sqrt{N}} \langle \mathcal{U}_N(t) \chi, \nielsNcal \mathcal{U}_N(t) \chi \rangle.
\end{equation}
Hence, to show the convergence of the many-body dynamics to the Gross-Pitaevskii equation, the central task is to bound the number of fluctuations $\langle \mathcal{U}_N(t) \chi, \nielsNcal \mathcal{U}_N(t) \chi \rangle$ uniformly in $N$. In the next section we explain how to obtain such a bound.

\section{Controlling the number of fluctuations}
In this section we use the following shorthands: $W_t := W(\sqrt{N} \varphi^{(N)}_t)$, $T_t := T(k_t)$, $ \langle . \rangle_t := \langle \mathcal{U}_N(t) \chi, .\,\,\mathcal{U}_N(t) \chi \rangle$.

We intend to use Gr\"onwall's lemma. Hence we compute the derivative
\begin{equation}
\label{nielsbeq:nexpectation}
\partial_t \langle \nielsNcal\rangle_t = \langle [i\nielsLcal_{N}(t),\nielsNcal]\rangle_t,\end{equation}                                                                                          
with $\nielsLcal_N(t)$ the time-dependent generator of $\mathcal{U}_N(t)$. Explicitly
\[
 \nielsLcal_N(t) = (i \partial_t T^*_t)T_t + T^*_t\big( (i \partial_t W^*_t)W_t + W^*_t \nielsHcal_N W_t \big) T_t =: (i \partial_t T^*_t)T_t + T^*_t \nielsLcal_N^{(0)}(t) T_t.
\]
The term $(i \partial_t T^*_t)T_t$ is harmless. Let us focus on the second term. In $\nielsLcal_N^{(0)}(t)$ we have
\[(i \partial_t W^*_t)W_t = -\sqrt{N} \int \nielsdi x\, a^*_x\, i\partial_t \varphi_t^{(N)}(x) + \nielshc\ (\,+\ \mathrm{irrelevant\   scalar}).\]
For $W^*_t \nielsHcal_N W_t$ we use \eqref{nielsbeq:weyl} and expand. We get summands which are linear in creation and annihilation operators and formally of order $N^{1/2}$; moreover quadratic summands of order one, cubics of order $N^{-1/2}$ and quartics of order $N^{-1}$.

Unlike in the mean-field regime \cite{RS}, where the Hartree equation implies complete cancellation of the linear terms in $W^*_t \nielsHcal_N W_t$ with $(i \partial_t W^*_t)W_t$, the modified Gross-Pitaevskii equation~\eqref{nielsbeq:modGP} leaves us with the linear, large remainder
\begin{equation}
 \label{nielsbeq:remainder}
N^{1/2} \int \nielsdi x \left( N^3 V(N.)\left(1-f\left(N.\right)\right) \ast \lvert \varphi^{(N)}_t \rvert^2 \right)\!(x)\,\varphi^{(N)}_t(x)\, a^*_x + \nielshc 
\end{equation}
The key observation is that by conjugating $\nielsLcal_N^{(0)}(t)$ with $T_t$, using \eqref{nielsbeq:bogtrafo} and expanding, we get (among many other terms) cubic terms which are not normal-ordered. Normal-ordering them by the canonical commutation relations gives rise to a linear term which cancels \eqref{nielsbeq:remainder}.

Similarly we get cancellations between quadratic and quartic terms of $\nielsLcal_N^{(0)}(t)$: we conjugate them with $T_t$ and expand the product; then normal-ordering of quartic terms gives rise to extra quadratic terms. Using the zero-energy scattering equation we now find a cancellation of some quadratic terms. (For identifying this cancellation think of $\sinh(k_t)(x;y)$ as $k_t(x;y)$ and of $\cosh(k_t)(x;y)$ as $\delta(x\!\,-\!\,y)$.)

These cancellations are crucial; they allow us to prove the operator inequality
\begin{equation}
  \label{nielsbeq:commutator}
 [i\nielsLcal_N(t),\nielsNcal] \leq \nielsHcal_N + C_t\left( \nielsNcal^2/N + \nielsNcal+1 \right). \nonumber
\end{equation}
(We denote by $C_t$ varying constants which may grow exponentially in $t$ (since we use bounds of the form $\nielsnorm{\varphi^{(N)}_t}_{H^n} \leq C e^{K \lvert t\rvert}$), but are independent of $N$.)

Next observe that $\nielsLcal_N(t) = \nielsHcal_N + \mathrm{other\ terms}$, where `other terms' can be bounded above and below by $\varepsilon \nielsHcal_N$ (any $\varepsilon >0$) and the number operator. Thus 
\begin{equation}
\label{nielsbeq:energy}
 \nielsHcal_N \leq C_t \left( \nielsLcal_N(t) + \nielsNcal^2/N + \nielsNcal+1 \right).  
\end{equation}
Thus we obtain $[i\nielsLcal_N(t),\nielsNcal] \leq C_t \left( \nielsLcal_N(t) + \nielsNcal^2/N + \nielsNcal+1 \right)$. It is possible to control $\langle \nielsNcal^2/N\rangle_t$ by $\langle (\nielsNcal+1)^2/N\rangle_{t=0}$ combined with $\langle\nielsNcal\rangle_t$.
This implies
\[\partial_t \langle \nielsNcal\rangle_t \leq C_t \langle \nielsNcal+1 + \nielsLcal_N(t) \rangle_t + C_t \langle (\nielsNcal+1)^2/N\rangle_0.\]
To close the scheme of Gr\"onwall's lemma, we need to control the growth of $\langle \nielsLcal_N(t)\rangle_t$.
Similar to the above estimates we find 
\begin{equation}
\label{nielsbeq:Ldot}
 \partial_t \langle \nielsLcal_N(t)\rangle_t = \langle \dot \nielsLcal_N(t)\rangle_t \leq C_t \langle \nielsLcal_N(t) + \nielsNcal+1\rangle_t + C_t \langle (\nielsNcal+1)^2/N\rangle_0.   \nonumber
\end{equation}
Combining the last two bounds, we obtain (for some fixed $D_t$ to be chosen later)
\[\partial_t \langle D_t(\nielsNcal+1) + \nielsLcal_N(t)\rangle_t  \leq C_t \langle D_t(\nielsNcal+1) + \nielsLcal_N(t) \rangle_t + C_t \langle (\nielsNcal+1)^2/N\rangle_0.\]
Thus, Gr\"onwall's lemma implies that for some $C, c>0$
\[\langle \nielsLcal_N(t) + D_t(\nielsNcal+1) \rangle_t \leq C\exp(c \exp(c t)) \langle \nielsLcal_N(0) + \nielsNcal+1 + \nielsNcal^2/N\rangle_{0}.\]
By \eqref{nielsbeq:energy} there exists a $C_t > 0$ such that $\nielsLcal_N(t) + C_t(\nielsNcal^2/N + \nielsNcal) \geq 0$. Choosing $D_t := C_t + 1$ we obtain
 \[\langle \nielsNcal\rangle_t \leq \langle \nielsLcal_N(t) + D_t(\nielsNcal^2/N + \nielsNcal)\rangle_t \leq C \exp(c \exp(c t)),\]
 which by \eqref{nielsbeq:numberoffluct} completes the proof of the main result.
\section*{Acknowledgments} The author would like to thank Benjamin Schlein and Gustavo de Oliveira for useful comments on a previous version of this review.
\bibliographystyle{alpha}
\bibliography{literature}

\begin{thebibliography}{ESY10}

\bibitem[BdS12]{BdS}
Niels Benedikter, Gustavo {de Oliveira}, and Benjamin Schlein.
\newblock {Quantitative Derivation of the Gross-Pitaevskii equation, to appear
  in \emph{Comm.\ Pure Appl.\ Math.}}
\newblock {\em Preprint arXiv:1208.0373}, 2012.

\bibitem[ESY10]{ESY2}
L\'aszl\'o Erd\H{o}s, Benjamin Schlein, and H.-T. Yau.
\newblock {Derivation of the Gross-Pitaevskii equation for the dynamics of
  Bose-Einstein condensate}.
\newblock {\em Ann.\ of Math.\ (2)}, 172(1):291--370, 2010.

\bibitem[LS02]{LS}
Elliot~H. Lieb and Robert Seiringer.
\newblock {Proof of Bose-Einstein condensation for dilute trapped gases}.
\newblock {\em Phys.\ Rev.\ Lett.}, 88:170409, 2002.

\bibitem[LSY00]{LSY}
Elliot~H. Lieb, Robert Seiringer, and Jakob Yngvason.
\newblock {Bosons in a trap: A rigorous derivation of the Gross-Pitaevskii
  energy functional}.
\newblock {\em Phys.\ Rev.\ A}, 61:043602, 2000.

\bibitem[Pic10]{Pickl}
Peter Pickl.
\newblock {Derivation of the Time Dependent Gross Pitaevskii Equation with
  External Fields}.
\newblock {\em Preprint arXiv:1001.4894}, 2010.

\bibitem[RS09]{RS}
Igor Rodnianski and Benjamin Schlein.
\newblock {Quantum fluctuations and rate of convergence towards mean field
  dynamics}.
\newblock {\em Comm.\ Math.\ Phys.}, 291(1):31--61, 2009.

\end{thebibliography}
\end{document}